\documentclass[11pt]{article}
\usepackage[latin1]{inputenc}
\usepackage[english]{babel}
\usepackage[namelimits]{amsmath}
\usepackage{authblk}
\usepackage{amssymb}
\usepackage{amsmath}
\usepackage{amsthm}

\begin{document}
\title{An algorithm to generate anisotropic rotating 
fluids with vanishing viscosity}
\author[1,2,3]{Stefano Viaggiu\thanks{s.viaggiu@unimarconi.it and viaggiu@axp.mat.uniroma2.it}}
\affil[1]{Dipartimento di Fisica Nucleare, Subnucleare e delle Radiazioni, Universit\'a degli Studi Guglielmo Marconi, Via Plinio 44, I-00193 Rome, Italy}
\affil[2]{Dipartimento di Matematica, Universit\`a di Roma ``Tor Vergata'', Via della Ricerca Scientifica, 1, I-00133 Roma, Italy.}
\affil[3]{INFN, Sezione di Napoli, Complesso Universitario di Monte S. Angelo,
	Via Cintia Edificio 6, 80126 Napoli, Italy.}
\date{\today}\maketitle
\begin{abstract}
Starting with generic stationary axially symmetric spacetimes depending
on two spacelike isotropic orthogonal coordinates $x^{1}, x^{2}$,
we build
anisotropic fluids with and without heat flow but with wanishing viscosity. In the first part of the paper, after 
applying
the transformation $x^1\rightarrow J(x^1)$,
 $x^2\rightarrow F(x^2)$(with $J(x^1), F(x^2)$ regular functions) 
to general metrics coefficients $g_{ab}(x^1,x^2)\rightarrow g_{ab}(J(x^1), F(x^2))$
with $G_{x^1 x^2}=0$, being $G_{ab}$ the Einstein's tensor,
we obtain that
${\tilde{G}}_{x^1 x^2}=0\rightarrow G_{x^1 x^2}(J(x^1),F(x^2))=0$. 
Therefore, the 
transformed spacetime is endowed with an energy-momentum tensor $T_{ab}$ with
expression $g_{ab}Q_{i}+heat\;term$ (where $g_{ab}$ is the metric and $\{Q_{i}\}, {i=1..4}$ are 
functions depending on the physical parameters of the fluid), i.e. without viscosity and generally with a non-vanishing 
heat flow. We show that after introducing suitable coordinates, we can obtain interior solutions that can be matched to the Kerr one on
spheroids or Cassinian ovals, providing the necessary mathematical machinery. 
In the second part of the paper we study the equation involving the heat flow and 
thus we generate anisotropic solutions with vanishing heat flow. In this frame, a class of asymptotically flat solutions with vanishing heat flow and viscosity can be obtained.
Finally, some explicit solutions are presented with possible applications to a string with
anisotropic source and a dark energy-like equation of state. 
\end{abstract}

\section{Introduction}
Exact metrics describing rotating, axially symmetric, isolated bodies are of
great astrophysical interest. Many methods have been developed in the 
literature \cite{1}-\cite{24} to build physically viable interior metrics. 
In particular, handble solutions describing a rotating body with
a perfect fluid source are not at our disposal, with the exception
of the Van Stockum solutions \cite{11} representing
pressureless dust spacetimes. Unfortunately, asymptotically flat 
solutions with a dust source are plagued by a curvature singularity at the 
origin of the polar coordinate system leading to a non-globally well  
defined mass function. Further, the technique named ``displace, cut, fill
and reflect'' has been applied to the van Stockum class of solutions
to build a rotating disk immersed in rotating dust (see \cite{13} and 
references therein). 
This method generates a distributional source of matter with 
rather unusual properties to describe ordinary galaxies.
In \cite{13}, the disk is obtained starting from Bonnor \cite{12}
dust solutions.
Although perfect fluids seem more appropriate to describe ordinary 
astrophysical objects,
anisotropic fluids (see \cite{a,m}) are finding applications in
physical situations where very compact objects come in action.\\ 
In the literature, only few solutions are present
depicting global physically reasonable sources suitable for
isolated rotating bodies. To this purpose, see 
the exception given in \cite{1,101} for a perfect fluid. Quite remarkably, recently \cite{25}
a procedure suitable for the static case has been proposed, and thus extended to the stationary rotating one
\cite{26,27}, in order to obtain a physically viable anisotropic interior source for the Kerr metric. This solution has interesting properties.
In particular, the solution in \cite{26} is equipped with a non-vanishing energy-momentum flux in the equatorial plane.
Inspired by the results in \cite{26}, we explore, from a mathematically point of view, 
the presence of viscosity and of a heat flow term in the equations governing axiallly-symmetric rotating 
anisotropic sources.\\
To this purpose, we show that, starting from a given stationary axially
symmetric seed spacetime (also an interior non-vacuum solution) with vanishing viscosity,
the map $x^1\rightarrow J(x^1)$,
$x^2\rightarrow F(x^2)$, applied to the metric functions $g_{ab}$, can be used to 
obtain global interior anisotropic fluids with heat flow and 
without viscosity,
(being $x^1, x^2$ two spacelike, orthogonal isotropic coordinates). 
Generally, the so generated solutions have a non-vanishing heat flow.
To obtain interior solutions with vanishing heat flow, a suitable 
procedure is also given. In this context,
a class of asymptotically flat solutions without heat flow 
and with mass-term is presented. 

In section 2 we present our technique to generate anisotropic fluids  
without viscosity and with heat flow with some examples. In section 3 we show how the procedure of section 2 can be used to obtain interior solutions matching the Kerr one on spheroids or 
Cassianian ovals boundary surfaces. In section 4 we study the technique to obtain metrics with vanishing heat flow.
In section 4 we also outline a simple 
procedure
to obtain asymptotically flat metrics representing anisotropic fluid with
vanishing heat flow and positive total mass and 
present a regular
class of solutions with a $G_3$ group of motion and containing the van Stockum metric \cite{11}
Section 5 collects some final remarks and conclusions.

\section{General case with vanishing viscosity}
The general expression for a metric
describing stationary axially symmetric spacetimes is given 
by \cite{20}
\begin{equation}
ds^2=g_{ab}dx^a dx^b\;,\; g_{ab}=- V_a V_b + W_a W_b + S_a S_b + L_a L_b,
\label{10}
\end{equation}
where ($x^1,x^2$) are spacelike orthogonal isotropic coordinates related to the
canonical Weyl coordinates ($\rho,z$) by 
${\rho}_{,x^1}=z_{,x^2},{\rho}_{,x^2}=-z_{,x^1}$, $x^3=\phi$ is the angular 
coordinate, $x^4=t$ is the time coordinate, and subindices with ``comma''
denote partial derivative and  
\begin{eqnarray}
& &V_a = \left(0, 0, \frac{N(x^1,x^2)}{\sqrt{f(x^1,x^2)}}, -\sqrt{f(x^1,x^2)}
\right),\nonumber\\
& &W_a= \left(0, 0, \frac{H(x^1,x^2)}{\sqrt{f(x^1,x^2)}}, 0\right),\nonumber\\
& &S_a = \left(e^{\frac{v(x^1,x^2)}{2}}, 0, 0, 0\right),\nonumber\\
& &L_a = \left(0, e^{\frac{v(x^1,x^2)}{2}}, 0, 0\right)\label{11}
\end{eqnarray}
an orthonormal basis.
For $H(x^1,x^2)=\rho(x^1,x^2)$ the metric is expressed in 
the so called Papapetrou gauge. It is a simple matter to verify
that the tensor $T_{ab}$ generated via $G_{ab}$ by (\ref{10}) has the 
following non-zero components: $(tt), (\phi\phi), (x^1 x^1), (x^2 x^2), 
(t\phi), (x^1 x^2)$, i.e. generally we have an anisotropic source matter.
Ordinary stars are not composed of anisotropic matter: it is generally accepted that 
perfect fluids can depict ordinary stars. However, anisotropic fluids are raising increasing interest in the literature.
Anisotropic fluids are expected to arise for very compact objects such as
neutron stars, where the suitable equation of state is a matter of debate.
For example, in \cite{27a} it is shown that anisotropies can modify the physical parameters of the fluid
and also the critical mass, the stability and the redshift of the stars. Moreover, anisotropic fluids are used to study
boson stars, where anisotropic stresses are naturally required \cite{27b}. It should also be noticed that the emission of 
gravitational waves from a given source is dependent on its equation of state, and thus the study of anisotropic 
energy-momentum tensors can be of great astrophysical interest in order to have possible hints on the frequency's emission of gravitational waves allowed in a general context. The use of anisotropic fluids is thus important for rotating fluids. 
This is in part also motivated by the fact that, thanks to unsolved formidable mathematical issues principally due to the complexity of the equations involved, no physically reasonable perfect fluid sources for realistic rotating isolated objects
have been found at present day. 
Anisotropic fluids are thus a natural arena to study the effects of rotation in general relativity, in particular when a huge
rotation is expected to modify the equation of state of non-rotating or slowly rotating objects. Dissipative effects are also expected, for example, in rotating neutron stars, where convective modes come into action.
Moreover, anisotropic fluids can arise from magnetic fields and collisionless
relativistic particles thanks to gravitational waves emission.
Concerning the most general energy 
momentum tensor generated by (\ref{10}) we have:
\begin{eqnarray}
& & T_{ab}=EV_a V_b+P_{x^1}S_a S_b+P_{x^2} L_a L_b+P_{\phi} W_a W_b+\label{q1}\\
& & +K\left(V_a W_b+W_a V_b\right)+\mathcal{H}\left(S_a L_b+L_a S_b\right),\nonumber
\end{eqnarray}
where $E$ is the energy density, $\{P_{x^1}, P_{x^2}, P_{\phi}\}$ the principal stresses, $K$ the heat flow parameter and 
$\mathcal{H}$ the viscosity one. Parameter $K$ depicts heat transfer in a given space equipped with a non-uniform
temperature distribution. In our frame, it is associated to the basis $V_a$ and $W_a$ and thus it is related, thanks to the axisymmetry of our background metric (\ref{10}), to transfer heat along the rotation axis. This effect is also present in the 
interior Kerr solution found in \cite{26,27}.\\
Concerning the viscosity parameter $\mathcal{H}$, as well known, it is not simple to treat \cite{27c} and it
is expected to depict peculiar dissipative properties of the fluid. In practice, viscosity depicts
internal friction due to a relative velocity between two adjacent
elements of the fluid. In particular, for the background (\ref{10}), the friction is expected to be on the $x^1-x^2$ plane.
Hence, from a physical point of view, the request of a vanishing viscosity practically implies that no internal friction is present 
in the $x^1-x^2$ plane and the only dissipative effect is the one due to the transfer heat along the rotation axis. This is an
assumption that is expected to hold, for example, in ordinary situations where magnetic fields play a non-relevant 
role. From a mathematical point of view, the vanishing of $\mathcal{H}$ is related to the vanishing of the 
$T_{x^1 x^2}$ component of the energy momentum tensor. For $G_{x^1 x^2}$ we found
\begin{eqnarray}
& &G_{x^1 x^2}=\frac{1}{2 f^2 H^2}A(x^1,x^2),\nonumber\\
& &A(x^1,x^2)=f_{,x^2}f_{,x^1}H^2+2f^2HH_{,x^1,x^2}+ff_{,x^2}NN_{,x^1}+
ff_{,x^1}NN_{,x^2}-\nonumber\\
& &ff_{,x^1}HH_{,x^2}-v_{,x^1}f^2HH_{,x^2}-v_{,x^2}f^2HH_{,x^1}-
f_{,x^2}f_{,x^1}N^2-\nonumber\\
& &ff_{,x^2}HH_{,x^1}-f^2N_{,x^2}N_{,x^1}.\label{12}
\end{eqnarray} 
Our technique is the following. Suppose to start with a generic line element (\ref{10}):
by applying the map $x^1\rightarrow J(x^1)$, 
$x^2\rightarrow F(x^2)$ to the metric functions $g_{ab}$ only, we generate a new spacetime with line element given by:
\begin{equation}
ds^2={\tilde{g}}_{ab}dx^a dx^b\,
\label{a10}
\end{equation}
where ${\tilde{g}}_{ab}={\tilde{g}}_{ab}(J(x^1),F(x^2))$ with the coordinates used in (\ref{10}) left unchanged 
\footnote{It is worth to being noticed that the map is applied to the metric functions $g_{ab}$ and not to the line element $ds^2=g_{ab}dx^a dx^b$, i.e. it is not obviously a coordinate transformation.}.
With this map, the (\ref{12}) 
is transformed to ${\tilde{G}}_{x^1 x^2}$:
\begin{eqnarray}
& & G_{x^1 x^2}\rightarrow {\tilde{G}}_{x^1, x^2}=\frac{1}
{2{\tilde{H}}^2{\tilde{f}}^2}F_{,x^2}J_{,x^1}
{\tilde{A}}(J(x^1),J(x^2)),\label{13}\\
& &{\tilde{A}}(J(x^1),F(x^2))=A((x^1\rightarrow J(x^1)), (x^2\rightarrow F(x^2))),\label{113}
\end{eqnarray}
where in (\ref{113}) the
partial derivatives with respect to $(x^1,x^2)$ are substituted
by the
partial derivatives with respect to $J,F$:
\begin{eqnarray}
& & {\tilde{A}}(J,F)={\tilde{f}}_{,F}{\tilde{f}}_{,J}{\tilde{H}}^2+
2{\tilde{f}}^2{\tilde{H}}{\tilde{H}}_{,F,J}+{\tilde{f}}{\tilde{f}}_{,F}{\tilde{N}}{\tilde{N}}_{,J}+
{\tilde{f}}{\tilde{f}}_{,J}{\tilde{N}}{\tilde{N}}_{,F}-\nonumber\\
& & {\tilde{f}}{\tilde{f}}_{,J}{\tilde{H}}{\tilde{H}}_{,F}-{\tilde{v}}_{,J}
{\tilde{f}}^2{\tilde{H}}{\tilde{H}}_{,F}-{\tilde{v}}_{,F}f^2{\tilde{H}}{\tilde{H}}_{,J}-
{\tilde{f}}_{,F}{\tilde{f}}_{,J}{\tilde{N}}^2-\nonumber\\
& & {\tilde{f}}{\tilde{f}}_{,F}{\tilde{H}}{\tilde{H}}_{,J}-{\tilde{f}}^2{\tilde{N}}_{,F}{\tilde{N}}_{,J}.\label{a12}
\end{eqnarray}
In what follows, functions
marked with a "tilde" will be in agreement with notation (\ref{113}).
Therefore, starting with a generic seed metric with $G_{x^1 x^2}=0$,
the transformed solution has again 
${\tilde{G}}_{x^1 x^2}=0$, provided that ($x^1,x^2$) are spacelike othogonal
isotropic coordinates.
As a result, 
the energy-momentum tensor so generated can be written as
\begin{eqnarray}
& &{\tilde{T}}_{ab}=\overline{E} 
{\tilde{V}}_a {\tilde{V}}_b + {\overline{P}}_{x^1} 
{\tilde{S}}_a {\tilde{S}}_b + {\overline{P}}_{x^2} 
{\tilde{L}}_a {\tilde{L}}_b +\nonumber\\ 
& &{\overline{P}}_{\phi}{\tilde{W}}_a{\tilde{W}}_b +
\overline{K}({\tilde{V}}_a{\tilde{W}}_b + 
{\tilde{W}}_a{\tilde{V}}_b),\label{14}
\end{eqnarray} 
where $\overline{E}$ denotes the transformed energy-density of the source, 
${\overline{P}}_{x^1},{\overline{P}}_{x^2},{\overline{P}}_{\phi}$ the transformed
principal stresses, and $\overline{K}$ the heat flow.\\
For the heat flow obtained from (\ref{14}), we obtain
\begin{equation}
\overline{K}=\frac{e^{-\tilde{v}}}{2}\left[
\frac{{\tilde{N}}_{,\alpha,\alpha}}{\tilde{H}}-
\frac{{\tilde{N}}_{,\alpha}
{\tilde{H}}_{,\alpha}}{{\tilde{H}}^2}-
\frac{\tilde{N}}{{\tilde{f}}{\tilde{H}}}
{\tilde{f}}_{,\alpha,\alpha} + 
\frac{\tilde{N}}{{\tilde{f}}
{\tilde{H}}^2}
{\tilde{f}}_{,\alpha}{\tilde{H}}_{,\alpha}\right],\label{15}
\end{equation}
where a summation with respect to $\alpha =(x^1,x^2)$ is implicit.
Equation (\ref{15}) will be discussed in the section 4.\\
Another equation of interest is the following
\begin{equation}
{\tilde{G}}_{x^1 x^1}+
{\tilde{G}}_{x^2 x^2}=-e^{\tilde{v}}
\left({\overline{P}}_{\rho}+{\overline{P}}_z\right)=
-\frac{{\tilde{H}}_{,\alpha,\alpha}}{\tilde{H}}.
\label{16}
\end{equation}
Equation (\ref{16}) gives us information about the equation of state
of the source of the transformed solution. In fact, if 
$\tilde{H}(x^1,x^2)$ is an harmonic
function (${\tilde{H}}_{,\alpha,\alpha}=0$), and 
if $({\tilde{P}}_{x^1},{\tilde{P}}_{x^2})\neq 0$, 
then the generated solution has equation of state
${\overline{P}}_{x^1}=-{\overline{P}}_{x^2}$: 
this is the case of generating solutions
with $\tilde{H}=\rho(x^1,x^2)$ (isotropic coordinates). 
When ${\overline{P}}_{x^1}={\tilde{P}}_{x^2}=0$, 
the generated solution can be a perfect fluid with a dust 
source or an anisotropic
fluid with string tension ${\overline{P}}_{\phi}$. 
An example of a class of solutions
containing both these equation of state will be given at the end of 
section 4.\\
If we want to consider more general situations, we have to
choose a non-harmonic function for $\tilde{H}$. Thus, if we start with a seed
metric with $H=\rho(x^1, x^2)$, i.e. a harmonic expression with 
${\rho}_{,\alpha,\alpha}=0$
, then by taking  $J(x^1)\neq x^1,F(x^2)\neq x^2$, we can build solutions with 
${\overline{P}}_{x^1}\neq -{\overline{P}}_{x^2}$. 
In summary, if we want to build
perfect fluid sources, then a necessary but (obviously) 
not sufficient condition is that
$\tilde{H}$ is not a harmonic function. 
This consideration means that our method
is also compatible with the generation of perfect 
fluid sources with non-vanishing
hydrostatic pressure.\\
Concerning the eigenvalues of (\ref{14}), i.e. $||T_{ab}-\lambda g_{ab}||=0$,
we have ${\lambda}_{\rho}=P_{\rho}, {\lambda}_{z}=P_{z}$ and with
the eigenvalues ${\lambda}_{t,\phi}$ that can be complex conjugate.
If $K=0$, then ${\lambda}_{t}=-E, {\lambda}_{\phi}=P_{\phi}$.\\

\subsection{An example: dust seed source}
We use canonical Weyl coordinates.
The line element appropriate for a stationary axially-symmetric dust
source is 
\begin{equation}
ds^2= e^{v}[d{\rho}^2+dz^2]+{\rho}^2 d{\phi}^2-
  {(dt-Nd\phi)}^2,
\label{1}
\end{equation}
where $N$ and $v$ depend on the canonical Weyl coordinates ($\rho,z$).
For (\ref{1}), Einstein's equations $G_{ab}=-T_{ab}$ give
\begin{eqnarray}
& &{\tilde{\nabla}}^2 N = 0\;,\;E=\frac{e^{-v}N_{,\alpha}^2}{{\rho}^2}\;,\;
{\tilde{\nabla}}^2={\partial}_{,\alpha,\alpha}-\frac{1}{\rho}
{\partial}_{,\rho},\nonumber\\
& &v_{,\rho}=\frac{N_{,z}^2-N_{,\rho}^2}{2\rho}\;\;,\;\;
   v_{,z}=-\frac{N_{,\rho}N_{,z}}{\rho}, \label{2}
\end{eqnarray}
In particular, the component
$(\rho z)$ is
\begin{equation}
G_{\rho z}=-\frac{1}{2{\rho}^2}\left[N_{,\rho}N_{,z}+\rho v_{,z}\right]
\label{3}
\end{equation}
that, for dust sources, is vanishing. In \cite{13}, the map $J(\rho)=\rho$,
$z\rightarrow |z|$ has been used to generate a disk immersed in a cloud
of dust. Generally, 
after applying the map $\rho\rightarrow J(\rho)$, $z\rightarrow F(z)$, 
for the component $(\rho z)$ we have
\begin{equation}
{\tilde{G}}_{\rho z}=-\frac{1}{2J^2}F_{,z}J_{,\rho}
\left[{\tilde{N}}_{,F}{\tilde{N}}_{,J}+
     J {\tilde{v}}_{,F}\right]=0.
\label{4}
\end{equation} 
Therefore, the energy-momentum tensor of the 
so obtained metric can be 
written as (\ref{14}) with
\begin{eqnarray}
& &{\tilde{V}}_a =\left(0, 0, \tilde{N}, -1 \right),\nonumber\\
& &{\tilde{W}}_a =\left(0, 0, J, 0\right),\nonumber\\
& &{\tilde{S}}_a =\left(e^{\frac{\tilde{v}}{2}}, 0, 0, 0\right),\nonumber\\
& &{\tilde{L}}_a =\left(0, e^{\frac{\tilde{v}}{2}}, 0, 0\right).\label{6}
\end{eqnarray}
Regarding the heat flow, we get
\begin{equation}
\overline{K}=\frac{e^{-\tilde{v}}}{2J}\left[{\tilde{N}}_{,\alpha,\alpha}-
  \frac{{\tilde{N}}_{,\rho}}{J}J_{,\rho}\right].
\label{b1}
\end{equation}
We consider, as a simple example, the seed metric given by
$N=c{\rho}^2 z, v=c^2{\rho}^4/8-c^2{\rho}^2 z^2$.
Setting $J(\rho)=\rho, F(z)=bz$, with $b$ a constant, 
for the generated solution we have the expression:
\begin{eqnarray}
& &\tilde{N}
=cbz{\rho}^2\;,\;
\tilde{v}=\frac{c^2{\rho}^4}{8}-c^2 b^2{\rho}^2 z^2,\nonumber\\
& &{\overline{P}}_{\rho}
=-{\overline{P}}_z=e^{-\tilde{v}}c^2{\rho}^2(1-b^2)\;,\;
{\overline{P}}_{\phi}=\frac{3}{4}c^2 e^{-\tilde{v}}{\rho}^2(1-b^2),\nonumber\\
& &{\overline{E}}=\frac{c^2}{4}e^{-\tilde{v}}\left[16 b^2z^2+7b^2{\rho}^2
-3{\rho}^2\right].
\label{8}
\end{eqnarray}
Solution (\ref{8}) is not asymptotically flat, but is regular and satisfies
all energy conditions for $|b|\geq 1$. For $b^2=1$, dust solution is
regained. Note that, for all the generated solutions, we have 
${\tilde{G}}_{\rho\rho}+{\tilde{G}}_{zz}
=-\frac{{J}_{,\rho,\rho}}{J}$, 
that, for $J=\rho$, leads to 
${\overline{P}}_{\rho}=-{\overline{P}}_z$.\\ 
Incidentally, solution (\ref{8}) has vanishing heat flow.

Another class of seed metrics are the ones 
asymptotically flat. For example, we can take the 
dipole Bonnor metric \cite{12} $N=c{\rho}^2{({\rho}^2+z^2)}^{-3/2}$
and perform a suitable map. Starting from an asymptotically
flat solution, sufficient conditions to obtain asymptotically flat and regular
metrics are:
\begin{eqnarray}
& &\lim_{z \to \pm\infty}F(z)=+\infty\;,\;F(z)\neq 0\;\;\;
\forall z \in \mathbf{R},\label{19}\\
& & for\;\; |z|\rightarrow\infty\;,\;F(z)=|z|+o(1)\nonumber\\
& &for\;\;\rho\rightarrow 0\;,\;J(\rho)=\rho+o(1)\;\;,\;\;
at \;\; infinity \;\;J(\rho)=\rho+o(1)\nonumber.
\end{eqnarray}
To satisfy conditions (\ref{19}), starting with the dipole Bonnor solution,
the simplest map we can take is  
$J(\rho)=\rho$,
$F(z)=\sqrt{z^2+b^2}$, with $b$ a constant. Note that by setting $b=0$ 
we are within the method depicted in \cite{13}.

\section{Matching the Kerr metric on Cassinian ovals: mathematical machinery}

As stated above, the map of section 2 does apply to isotropic harmonic coordinates $x^1,x^2$ with
$\Delta\rho(x^1,x^2)=0, \Delta={\partial}^2_{x^1}+{\partial}^2_{x^2}$.
In this section we show how to apply the procedure depicted in section 2 to obtain interior solutions that can be matched on a suitable surface to the Kerr metric.\\
The Kerr metric  with ADM mass $M$ and angular momentum for unit mass $a$
can be expressed in the Papapetrou gauge by using  (with $a^2<M^2$ black hole solution \cite{28}) 
spheroidal prolate coordinates $\mu,{\overline{\theta}}$ with
$\rho = \sinh\mu\sin{\overline{\theta}}\;,\;z=\cosh\mu\cos{\overline{\theta}}$:
\begin{equation}
ds^2=f^{-1}(\mu,{\overline{\theta}})\left[e^{2\gamma(\mu,{\overline{\theta}})}(d{\mu}^2+
d{{\overline{\theta}}}^2)+{\rho}^2d{\phi}^2\right]-
f(\mu,{\overline{\theta}}){\left[dt-\omega(\mu,{\overline{\theta}}) d\phi\right]}^2,
\label{B1}
\end{equation}
and
\begin{eqnarray}
& & f=\frac{p^2{\cosh}^2\mu+q^2{\cos}^2{\overline{\theta}}-1}
{{\left(p\cosh\mu+1\right)}^2+q^2{\cos}^2{\overline{\theta}}}\;,\; 
\omega=2\frac{q}{p}\frac{(p\cosh\mu+1){\sin}^2{\overline{\theta}}}
{\left[p^2{\cosh}^2\mu-1+q^2{\cos}^2{\overline{\theta}}\right]}, \nonumber \\
& & e^{2\gamma} ={\left(p^2{\cosh}^2\mu-1+q^2{\cos}^2{\overline{\theta}}\right)},
\label{B2}
\end{eqnarray}
and $p=1/M$, $q=a/M$, with $M^2-a^2=1$.
We can apply, to the metric (\ref{B1}) the map ${\overline{\theta}}\rightarrow F({\overline{\theta}})={\overline{\theta}}$ 
and $\mu\rightarrow J(\mu)$ where $J(\mu=k)=\mu$ for 
$k\in\mathbf{R^+}$. In this way we can regularize the interior solution and perform a matching on some spheroidal closed surface 
$\mu=k$. However, this surface is prolate and physical intuition suggests an oblate surface of rotation rather than a prolate one.
In the mainstream present in the literature, the matching with the Kerr solution is performed on spheroids (prolate or oblate), but, for example, no perfect fluid interior solution has been found. It is thus evident the necessity to explore more general surfaces that could be suitable
with an interior perfect fluid solution. In what follows we give the mathematical machinery to match the Kerr solution on general 
surfaces with the help of the technique of section 2.\\
To start we fix a coordinate system $x^1,x^2,\phi,t$ with the line element in the Papapetrou form (\ref{B2}) together with the 
harmonic condition $\Delta{\rho}(x^1,x^2)=0$. For practical purposes \cite{18}, 
more easy computations can be done with boundary non-null surfaces $S$
with equation $T(x^{\alpha})=0$ with $x^{\alpha}$ assumed to be $x^1$ or $x^2$ and with unit normal 
$n_{\alpha}=\pm({\partial}_{\alpha}T/\sqrt{{\partial}_{\beta}T{\partial}^{\beta}T})$. The standard procedure \cite{18} is to impose
the continuity of the first fundamental form (the pull-back of the metric on $S$) and the second fundamental form $K_{ij}$
(with $K_{ij}=n_{i;j}$, with "$;$" as the covariant derivative on $S$). It is a simple matter to verify that a sufficient (but not necessary)
way to satisfy the conditions above is to choose the metric coefficients $g_{\alpha\beta}$ as $C^1$ functions on $S$.
The next step is to identify a sufficiently general and reasonable class of matching surfaces. To this purposes, in \cite{30} it has been shown
that, within the well known Ehler method, interior rotating solutions with an energy density constant on Cassinian's ovals 
quartic surfaces can be obtained. Cassinian  ovals have been introduced by Giovanni Domenico Cassini in 1680 in order to
substitute the ellipses with ovals to describe planets orbits. In Newtonian mechanics this task cannot be accomplished, but in general relativity, as shown in \cite{30}, Cassinian ovals can arise as a possible configuration surface suitable for rotating bodies. 
Hence, we could speculate that, until now, perfect fluid Kerr interior metrics
has not been obtained because non-suitable surfaces have been chosen as 
possible ones for rotating bodies. To this purpose, Cassinian ovals are defined as the set of points $P$ such that the product of the distances between $P$ and two fixed points $F_1=(-\ell,0)$ and $F_2=(\ell,0)$ called foci is constant. In order to obtain coordinates suitable for rotating surfaces, we write down the Cassinian equation using Weyl canonical coordinates $\rho,z$:
\begin{equation}
{({\rho}^2+z^2-{\ell}^2)}^2+4{\ell}^2 z^2=4{\ell}^2 c^2,
\label{B3}
\end{equation}
with $c$ as a constant: for $c<\ell/2$ we have toroidal configurations, while for $\ell\leq c$ we have ellipsoidal-like (quartic)
surfaces. We need harmonic spatial coordinates $\{m,\theta\}$ such that the boundary surface is obtained at $m=k\in\mathbf{R}$.
First of all, we define $t={({\rho}^2+z^2-{\ell}^2)}^2+4{\ell}^2 z^2$ with $t\in[0, \infty)$. To obtain harmonic coordinates, we introduce the
complex plan $\zeta$ with
\begin{equation}
\zeta={(\rho+\imath z)}^2-{\ell}^2=u+\imath v,\;\;\;u={\rho}^2-z^2-{\ell}^2,\;\;v=2z\rho,
\label{B4}	
\end{equation}	
together with:
\begin{eqnarray}
& & \zeta=\sqrt{t}(\cos\theta+\imath\sin\theta),\label{B5}\\
& &\cos\theta=\frac{{\rho}^2-z^2-{\ell}^2}{\sqrt{{({\rho}^2+z^2-{\ell}^2)}^2+4{\ell}^2 z^2}}.\nonumber
\end{eqnarray}
We obtain a polar representation of the coordinates $u,v$ by setting
$u=e^m\cos\theta,\;v=e^m\sin\theta$ with $\sqrt{t}=e^m$ ($m\in(-\infty,\infty)$). 
Finally we obtain the relations between the Weyl coordinates 
$\rho,z$ and the harmonic Cassinian's ones $m,\theta$:
\begin{eqnarray}
& &\rho=\frac{Q}{\sqrt{2}},\;\;\;\;z=\frac{v}{Q\sqrt{2}},\label{B6}\\
& & Q=\sqrt{u+{\ell}^2+\sqrt{{(u+{\ell}^2)}^2+v^2}}.\nonumber 
\end{eqnarray}
With Cassinian harmonic coordinates $m,\theta$ depicting at $m=k$ Cassinian ovals, the Kerr metric can be expressed in 
terms of these coordinates\footnote{This can be done expressing the Kerr metric in the Weyl coordinates and thus perform the
eq.(\ref{B6}.)}. However, in this paper we are only interested in presenting a further application of the algorithm in section 2.
To this purpose, we only need to observe that in the coordinates $\mu,\theta$, the line element (\ref{B1}) becomes:
\begin{equation}
ds^2=f^{-1}(m,\theta)\left[e^{2\gamma(m,\theta)}(d{m}^2+d{\theta}^2)+{\rho}^2(m,\theta)d{\phi}^2\right]-
f(m,\theta){\left[dt-\omega(m,\theta) d\phi\right]}^2.
\label{B7}
\end{equation}
To fulfill matching conditions on Cassinian ovals, the simplest map we can consider is:
\begin{equation}
\theta\rightarrow F(\theta)=\theta,\;\;\;\;m\rightarrow J(m).
\label{B8}
\end{equation}
With the map (\ref{B8}), $\theta$ is left unchanged.\\
Starting with 
the metric (\ref{B7}) with $\Delta\rho(m,\theta)=0$, the map (\ref{B8}) will transform 
$\rho(m,\theta)\rightarrow H(m,\theta)\neq\rho(m,\theta)$ with $H_{,m,m}+H_{,\theta,\theta}\neq 0$ and 
as a consequence $H$ is no longer a harmonic function. Thanks to equation (\ref{16}), this does imply that
$P_m+P_{\theta}\neq 0$ and more general equations of state than the ones obtained in the examples of section 2 can be obtained, leaving open 
the possibility to obtain perfect fluid sources. From a mathematical point of view \cite{31}, $\rho$ in (\ref{B2}) or (\ref{B7}) 
is nothing else but the determinant of the $2-$metric $g_2$ spanned by the Killing vectors ${\partial}_t$ and
${\partial}_{\phi}$ and characterizes a measure of the orbits of the isometry group \cite{31}. Hence, in order to obtain an interior source from a vacuum solution with more general equations of state than the (\ref{16}), the measure of the orbits of the isometry group must be changed.\\
As a final step, we must specify $J(m)$ in (\ref{B8}), with the conditions: $\forall m\in(-\infty, k]\;J(m)\in C^1$ and
\footnote{Since $\sqrt{t}=e^m$ this does imply that the matching surface $S$ is defined by $t=e^{2k}$.} $J(m=k)=m$. 
Obviously, these conditions can be easily fulfilled. As an example, we can take
\begin{eqnarray}
& & J(m)=m+w{(m-k)}^b,\;w\in\mathbf{R^+},\;b>1,\label{B9}\\
& & J(m)=m+\sum_{i=1}^n w_i{(m-k)}^{i+\epsilon},\;\;w_i\in\mathbf{R^+},\;\;\epsilon>0.\label{B10}
\end{eqnarray}
The interior Kerr solutions generated with the maps (\ref{B9}) and (\ref{B10}) are regular and smoothly match the vacuum Kerr solution
at the Cassinian surfaces $m=k$. This certainly will be matter for futures calculations. We stress that, with our procedure, 
the finding of physically and mathematically reasonable (perfect fluid ?) interior rotating solutions for real astrophysical objects
must require a reasonable matching surface: Cassinian ovals can offer a possible realistic solution for this issue.   

\section{Vanishing Heat Flow}   
In this second part of the paper, we study the equation governing heat flow term. 
To the solutions generated with the technique of this section, we can apply the transformation
$x^1\rightarrow J(x^1)$, $x^2\rightarrow F(x^2)$ and thus we can generate new solutions with vanishing viscosity and 
generally with a non-zero heat flow term.\\
The equation $K=0$ is nothing else but
\begin{equation}
N_{,\alpha,\alpha}-\frac{1}{H}N_{,\alpha}H_{,\alpha}-
\frac{N}{f}\left(f_{,\alpha,\alpha}-
\frac{1}{H}f_{,\alpha}H_{,\alpha}\right)=0.
\label{A1}
\end{equation}
Equation ({\ref{A1}) can be satisfied by taking, for example $N=f$.
Generally, if we take a solution $N(x^1,x^2), f(x^1,x^2)$ of (\ref{A1}), then
also $N(G,T),f(G,T)$ is, being 
$G(x^1,x^2)$ an arbitrary harmonic function and $T(x^1,x^2)$ its harmonic
conjugate (see \cite{23}).\\
By posing $\omega=\frac{N}{f}$ and introducing the Ernst-like potential 
$\Phi$ such that
\begin{equation}
{\Phi}_{,x^1}=\frac{f^2}{H}{\omega}_{,x^2}\;\;,\;\;
{\Phi}_{,x^2}=-\frac{f^2}{H}{\omega}_{,x^1},
\label{f1}
\end{equation}
we see that the integrability condition ${\Phi}_{,x^1,x^2}={\Phi}_{,x^2,x^1}$ 
for (\ref{f1}) leads exactly to equation (\ref{A1}). Further,
by taking the integrability condition for (\ref{f1}) in terms of $\Phi$ 
(i.e. ${\omega}_{,x^1,x^2}={\omega}_{,x^2,x^1}$), we obtain:
\begin{equation}
f\left[{\Phi}_{,\alpha,\alpha}+\frac{H_{,\alpha}}{H}{\Phi}_{,\alpha}\right]-
2{\Phi}_{,\alpha}f_{,\alpha}=0.
\label{f2}
\end{equation} 
By setting $H=\rho(x^1,x^2)$, equation (\ref{f2}) is one of the 
two equations of the Ernst method for the vacuum
expressed in the Papapetrou gauge, the other being
\begin{equation}
f\left[f_{,\alpha,\alpha}+\frac{{H}_{,\alpha}}{H}{f}_{,\alpha}\right]+
{\Phi}^{2}_{,\alpha}-f^{2}_{,\alpha}=0.
\label{f3}
\end{equation}
For $H=\rho$, (\ref{f2}) and (\ref{f3}) are the Ernst
equations expressed in the Papapetrou gauge.
Therefore, considering both (\ref{f2}) and 
(\ref{f3}), we have the 
vacuum Ernst equations, while the equation (\ref{f2}) without 
(\ref{f3}) is compatible with spacetimes with anisotropic 
pressure and vanishing heat flow and viscosity.\\
In order to obtain explicit solutions, we can for example separate the (\ref{A1}) by posing
\begin{equation}
N_{,\alpha,\alpha}-\frac{1}{H}N_{,\alpha}H_{,\alpha}=0\;,\;
f_{,\alpha,\alpha}-\frac{1}{H}f_{,\alpha}H_{,\alpha}=0,
\label{A2}
\end{equation}
that, after setting $H(x^1,x^2)=\rho$, becomes ${\tilde{\nabla}}^2 N=0,
{\tilde{\nabla}}^2 f=0$, that for $f=1$ reduces to the dust case.
The next step is to consider the equation $G_{x^1 x^2}=0$, with $G_{x^1 x^2}$
given by (\ref{12}). This equation involves
both $v_{,x^1}$ and $v_{,x^2}$ and therefore can be integrated directly 
by imposing the integrability condition $v_{,x^1, x^2}=v_{,x^2,x^1}$ 
(see \cite{23}). However, if we take, for example, $H(x^1,x^2)=H(x^1)$, 
the term proportional to
$v_{,x^1}$ disappears in (\ref{12}), obtaining
\begin{eqnarray}
& &v_{,x^2}=\frac{1}{f^2HH_{,x^1}}B(x^1,x^2),\label{A3}\\
& &B(x^1,x^2)=f_{,x^2}f_{,x^1}(H^2-N^2)+ff_{,x^2}NN_{,x^1}+
ff_{,x^1}NN_{,x^2}-\nonumber\\
& &ff_{,x^2}HH_{,x^1}-f^2N_{,x^2}N_{,x^1}.\nonumber
\end{eqnarray} 
If we set $H(x^1,x^2)=H(x^2)$, then the term involving $v_{,x^2}$ disappears,
and therefore we can easily calculate $v_{,x^1}$.\\
Obviously, to the equation (\ref{A3}) we can apply the map of section 2.

\subsection{A class of asymptotically flat solutions}

In what follows, without loss of generality, we adopt cylindrical
coordinates with $x^1=\rho, x^2=z$.\\
By inspection of equations (\ref{A2}), we see that given a solution $N$ for the first equation, any combination
$f=x+yN$ with 
$\{x,y\}\in\mathbf{R}$ is a solution for the second of (\ref{A2}). 
After imposing that $N\rightarrow 0$ at spatial infinity, we can obtain
asymptotically flat solutions.\\
Hence, 
looking for asymptotically flat solutions without heat flow we can take, for example, for $N$ an asymptotically flat solution of the
first of (\ref{A2}) with 
$N\simeq\frac{1}{\sqrt{{\rho}^2+z^2}}+o(1)$ at
spatial infinity. For the metric function $f$, we could take 
a generic linear combination of solutions $N_a$ of the first of (\ref{A2}):
$f=1-k_a N_a$ (with $k_a$ constant coefficients).\\ 
As an example, we can take (Bonnor)
\begin{equation}
H=\rho,\;N=c{\rho}^2{({\rho}^2+z^2)}^{-3/2},\;N\sim\frac{1}{\sqrt{{\rho}^2+z^2}},\;f=1-kN
\label{IC}
\end{equation}
With the solutions (\ref{IC}), we can 
integrate equation (\ref{A3}) to calculate the metric coefficient
$v$. Generally, we obtain a very complicated expression for $v$, but with the correct asymptotic
behaviour ($v\rightarrow 0$ at spatial infinity) suitable for
asymptotically flat metrics, 
provided that the integration constant is chosen to be zero.\\
Obviously, to the solutions and the technique presented in this section, we can apply the map of section 2.
To this purpose, note that equation (\ref{A1}) (and (\ref{A2})) is not invariant
in form under the map considered in section 2 and as a result starting from 
a seed solution with vanishing heat flow, the map 
$x^1\rightarrow J(x^1)$, $x^2\rightarrow F(x^2)$ generally does not generate
a solution with vanishing heat flow, but rather with vanishing viscosity. Hence, we can apply the transformation
$x^1\rightarrow J(x^1)$, $x^2\rightarrow F(x^2)$ of section 2 to spacetimes generated  
in this section to obtain new solutions with vanishing viscosity and non-zero heat flow.
We can thus start with the Bonnor solution $N=\frac{1}{\sqrt{{\rho}^2+z^2}}$ with the map (as a simple example)
\begin{equation}
J(x^1)=\rho,\;\; F(x^2)=\sqrt{z^2+a^2}, a\in\mathbf{R},
\label{CS}
\end{equation} 
to obtain regular solutions with anisotropic pressures and vanishing viscosity.

\subsection{Solutions with a $G_3$ group of motion}
As a physically interesting subcase, we consider 
solutions with a $G_3$  
group of motion with cylindrical symmetry. In this case, retaining
the condition $f=1-kN$, 
equation (\ref{A1}) becomes
\begin{equation}
\frac{H_{,\rho}}{H}=\frac{N_{,\rho,\rho}}{N_{,\rho}}.
\label{A4}
\end{equation} 
Equation (\ref{A4}) is integrable. Moreover, equation $G_{\rho z}=0$ is identically
satisfied, and therefore solutions depend on the arbitrary metric
function $v(\rho)$.\\ 
As a title of example we consider the solution $H=\rho,
N=\frac{c}{2}{\rho}^2$. We get
\begin{eqnarray} 
& &P_{\rho}=-P_z=\frac{(4v_{,\rho}-4kc\rho-4{\rho}^2 kcv_{,\rho}+
{\rho}^4 k^2c^2v_{,\rho}+2c^2\rho)}
{2\rho e^{v}{(-2+kc{\rho}^2)}^2},\nonumber\\
& &E=\frac{-e^{-v}}{2{(-2+kc{\rho}^2)}^2}[-6c^2+12kc+
({\rho}^4 k^2c^2+4-4{\rho}^2 kc)v_{,\rho,\rho}],\nonumber\\
& &E+P_{\phi}=\frac{4ce^{-v}(c-2k)}{{(-2+kc{\rho}^2)}^2}.\label{A5}
\end{eqnarray}
Note that, also in this case, thanks to (\ref{16}), by taking
$H\neq \rho$, we can obtain a more general equation of state than the one 
in (\ref{A5}).\\
The regularity of (\ref{A5}) is fulfilled
if $kc\leq 0$, and if, for $\rho\rightarrow 0$,
$v(\rho)$ looks as follows: $v=-a^2{\rho}^n+o(1)$, being $a,n$ numbers 
with $n\geq 2$. Further, energy conditions are satisfied by setting
$v_{,\rho,\rho}\leq 0$: this is a sufficient but not necessary condition.
In the limiting case $k=0, v=-\frac{c^2{\rho}^2}{4}$ and
we regain the van Stockum solution \cite{11}.\\
As a further example, we can look for solutions with string tension
and vanishing radial pressure, i.e. with $P_{\rho}=P_z=0, P_{\phi}\geq 0$.
We have:
\begin{eqnarray}
& &v=\frac{2k-c}{k(2-kc{\rho}^2)}\;\;,P_{\rho}=P_z=0,\nonumber\\
& &E=e^{-v}8c\frac{2k-c}{{(-2+kc{\rho}^2)}^3}\;\;,\;\;
P_{\phi}=e^{-v}4c^2k\frac{-2k+c}{{(-2+kc{\rho}^2)}^3}.\label{f5}
\end{eqnarray}
Solution (\ref{f5}) is regular for $c\geq 0, k\leq 0$ and satisfies
the weak and the strong energy conditions, but not the dominant
energy condition. Note that solution (\ref{f5}),
in the limit $\rho\rightarrow\infty $, has vanishing
$E, P_{\phi}$ and $v$.

As a final study, we can compare the features of the internal solutions generated in this paper with the Kerr interior one in terms of Cassinian ovals of section 3. Summarizing in a list we have:
\begin{itemize}
\item Kerr interior solution (\ref{B7})-(\ref{B10}): Vanishing viscosity; non-vanishing heat flow; no curvature singularities;
equation of state $P_m+P_{\theta}\neq 0$; matching smoothly to Kerr on Cassinian ovals.
\item Solution (\ref{8}): Vanishing viscosity; vanishing heat flow; no curvature singularities; string-like equation of state 
$P_{x^1}+P_{x^2}=0$; non-asymptotically flat; no matching with Kerr exterior solution.
\item Solutions (\ref{A2}) with (\ref{IC}): Vanishing viscosity; vanishing heat flow; curvature singularity at
$(\rho,z)=(0,0)$; equation of state $P_{\rho}+P_z=0$; asymptotically flat; no matching with the Kerr metric.
\item Solution (\ref{A2}) with the map (\ref{CS});
Vanishing viscosity; non-vanishing heat flow; no curvature singularities;
equation of state $P_{\rho}+P_z=0$; asymptotically flat; no matching with the Kerr metric.
\item Solution (\ref{A4})-(\ref{A5}): Vanishing viscosity; vanishing heat flow; no curvature singularities; equation of state
$P_{\rho}+P_z=0$; non-asymptotically flat; no matching with the Kerr metric.
\end{itemize}

\section{Conclusions and final remarks}
In this paper, we have presented methods to generate anisotropic fluids. Anisotropic sources are
of great astrophysical interest, for example in the context of very compact objects
at high densities.
Many generating methods are present
in the literature to obtain rotating fluid solutions. The most famous one is the Ehlers method \cite{k1}
where new stationary exterior solutions and interior ones with a one-parameter family
are obtained starting from static vacuum solutions. Within this
beautiful method, generally one generates solutions with curvature singularities. In \cite{c} Geroch showed that it is possible to obtain an infinite-parameter family of solutions. In \cite{k2} a technique has been presented for generating a two parameter family starting from vacuum solutions. With these techniques generally it is not easy to have a sound physical control of the so generated
energy-momentum tensor when rotating sources are investigated \cite{4,15,k3,k4,k5,k7,k8,k88}, or the generating methods have a particular specific equation of state \cite{k5}. Moreover, in order to depict astrophysical objects, we need solutions
representing an isolated body or a solution that can be matched to the exterior Kerr one on a suitable surface of rotation \cite{4,15,k3,k4,k5,k7,k8}.\\
With the algorithms presented in this paper, we can build
global solutions with and without heat flow and vanishing viscosity. In particular, under a suitable choice of the seed metric,
we can also obtain global solutions that are also asymptotically flat.\\ 
In the first part of the paper, we have presented the 
transformation properties of a generic stationary
axially symmetric line element, representing a vacuum or fluid filled solution, under the map 
$x^1\rightarrow J(x^1)$, $x^2\rightarrow F(x^2)$. We have shown that, starting
with a seed metric with vasnishing viscosity,
$G_{x^1 x^2}=0$, the generated solution has again
${\tilde{G}}_{x^1 x^2}=0$. 
Generally, the source of the so obtained metric is composed
of anisotropic fluid with heat flow. The limit appropriate for thin disk is given by setting 
$F(z)=|z|$ in the Weyl coordinates. 
In the literature \cite{13}, the disk has been matched with the
exterior dust dipole Bonnor solution. As a consequence, due to the fact
that asymptotically flat dust solutions have not mass term,
in the solution so obtained
unavoidably will appear exotic matter with negative mass and total zero mass.
Therefore, we argue that, for the reasonings made in this paper, we 
can obtain more realistic solutions with positive mass term and asymptotically
flat, with a progress with
respect to the method in \cite{13} . 
It is also interesting to note that for the starting seed solution, we can also take a vacuum solution of the Ernst equations, where
any fluid is absent and obviously the viscosity is zero. In particular, after applying a suitable map to the Kerr metric, we can obtain 
regular asymptotically flat solutions representing fluids with vanishing viscosity, non-zero heat flow and, after setting
$J(x^1)=\rho$ , equation of state with $P_{\rho}=-P_z$.\\
Moreover, after taking $F(x^2)=const.$ (or $H(x^1)=const.$), 
we can build solutions with
at least a  $G_3$ group of motion.\\
In section 4 we have shown how, within the technique of section 3, we can 
easily build interior solutions that can be matched to the Kerr one on suitable boundary surfaces. In particular, we present the general framework to match the Kerr solution on Cassinian ovals that are very interesting surface suitable to describe a rotating star.\\
Summarizing, the following possible advantages with the procedures outlined in this paper with respect to the ones usually present in the literature are:
\begin{enumerate}
\item We can obtain anisotropic fluids from known solutions with a $T_{ab}$ with the specific feature to have a vanishing viscosity,
i.e. the friction for the fluid is absent in particular in the $x^1-x^2$ plane by the properties of the component $G_{x^1 x^2}$
analyzed in section 2.
\item By means of the function $H$, we can obtain, thanks to equation (\ref{16}), anisotropic fluids with equation of state 
more general than the string ones with $P_{x^1}=-P_{x^2}$.
\item As shown at the end of section 2, we can generate asymptotically flat solutions, starting from the Bonnor ones,
that are regular on the whole axis of rotation and potentially solving the zero-mass problem of this class of solutions.
\item In section 4 we showed that it is possible to generate asymptotically flat solutions with also vanishing heat flow and with generally some curvature singularity. 
Starting from these solutions and after applying the map of section 2, we can build asymptotically flat solutions with non vanishing
heat flow, but with vanishing viscosity and regular everywhere.
\item In section 3 it is argued that, thanks to relativistic effects due to the rotation, the shape of the source can be more general
than the one provided by spheroids. To this regard, more general coordinates, the Cassinian ones, are introduced that permit us to explore more general revolution surfaces than the spheroidal ones expected in a Newtonian context. To these new coordinates, the algorithm of section 2 can be applied to obtain interior Kerr solutions that are regular everywhere.
To the best of my knowledge, this line of research is absolutely new and is certainly matter for future investigations.
\end{enumerate}
As a final consideration, we discuss the algorithm outlined in this paper with similar approaches recently appeared in the literature.
Our algorithm to find interior rotating solutions and in particular the one that is capable to be smoothly matched to the Kerr one can 
be related to the interesting recent works present in \cite{25,26,27,k9}. The algorithm in the aforementioned papers is based on an
ansatz applied to the Kerr exterior solution, and calculated at the boundary surface representing a spheroid. Also, in these papers the starting point is the line element in the Weyl-Papapetrou form (\ref{10}), and the interior solution is obtained by guessing in a
suitable way some component of the metric coefficients in such a way that the static solution represents a perfect fluid.
In this way regular metrics, satisfying the energy conditions and representing anisotropic fluids with heat flow and
viscosity are obtained. In fact, the so generated solutions have the components $(t\phi)$ and $(x^1 x^2)$ of $T_{ab}$ non-vanishing.
Hence, the generated solutions in \cite{26,27} can be seen in light of our results as solutions generated from guessing 
components of the Kerr metric that do not preserve the condition $G_{x^1 x^2}=0$
assuring a vanishing viscosity. However, our technique and the one in \cite{26,27} show that exists a systematic way to guess
a stationary vacuum solution in order to obtain a generic anisotropic source with and without viscosity and heat flow.
We stress that, in order to obtain interior Kerr solutions, more general boundary surfaces can be considered. In practice,
general relativistic effects can lead to a modification of the usual spheroids as boundary surfaces: Cassinian ovals, for their properties as quartic curves, are the natural candidates.

\section*{Acknowledgements} 
This paper is dedicated to the memory of my friens and collegue Roberto Bergamini (1940-2003) who suggested
to me the idea to use Cassinian ovals as suitable boundary surfaces to match the Kerr metric.

\end{document}